\documentclass[preprint,12pt]{elsarticle}

\usepackage{amssymb}
\def\aap{A\&A }
\def\apj{ApJ }

\journal{New Astronomy}

\begin{document}

\begin{frontmatter}



\title{The $\eta$~Carin\ae\ optical 2009.0 event, a new ``eclipse-like'' phenomenon}


\author{
E.~Fern\'andez-Laj\'us$^{a,b,}$\footnote{eflajus@fcaglp.unlp.edu.ar},
%       \thanks{Member of the Carrera del Investigador Cient\'ifico, CONICET, Argentina},
        C.~Fari\~na$^{a,b}$,
        J.P.~Calder\'on$^{a}$,
        N.~Salerno$^{a}$,
        A.F.~Torres$^{a,b}$,
        M.A.~Schwartz$^{a}$,
        C.~von Essen$^{a}$,
        F.~Giudici$^{a}$,
       % \and
        F.A.~Bareilles$^{a}$
}

\address{
	$^{a}$Facultad de Ciencias Astron\'omicas y Geof\'{\i}sicas (FCAG)-\\
              Universidad Nacional de La Plata (UNLP)

	$^{b}$Instituto de Astrof\'{\i}sica de La Plata (CCT La Plata - CONICET), Argentina

}

\begin{abstract}
The periodic events occurring in $\eta$~Carin\ae\/ have been widely
monitored during the last three 5.5-year cycles. The last one 
recently occurred in January 2009 and more exhaustive observations
have been made at different wavelength ranges.
If these events are produced when the binary components approach
periastron, the timing and sampling of the photometric features can
provide more information about the geometry and physics of the system.
Thus, we continued with our ground-based optical photometric
campaign started in 2003 to record the behaviour of the
2009.0 event in detail. This time the observation program included a 
new telescope to obtain information from other photometric bands.
The daily monitoring consists of the acquisition of CCD images
through standard $UBVRI$ filters and a narrow H$\alpha$ passband.
The subsequent differential photometry includes the central region
of the object and the whole Homunculus nebula.
The results of our relative $UBVRI$H$\alpha$ photometry, performed
from November 2008 up to the end of March 2009, are presented in this 
work, which comprises the totality of the event. The initial rising branch,
the maximum, the dip to the minimum and the recovering rising phase 
strongly resemble a kind of eclipse.
All these features happened on time - according to that predicted -
although there are some photometric differences in comparison 
with the previous event.
We made a new determination of 2022.8 days for the period value
using the present and previous ``eclipse-like'' event data.
These results strongly support the binarity hypothesis for $\eta$~Car.
In this paper, the complete dataset with the photometry
of the 2009.0 event is provided to make it readily available for further
analysis.
\end{abstract}

\begin{keyword}
stars: individual ($\eta$~Carin\ae) \sep
stars: variables: general \sep 
stars: binaries: close \sep
techniques: photometric

97.30.Eh \sep 97.80.Fk \sep 97.60.-s \sep 

\end{keyword}

\end{frontmatter}


\section{Introduction}
$\eta$~Carin\ae\ is one of the most massive and luminous stars in the Milky Way,
and the brightest Luminous Blue Variable star in the sky ($V$\,$\sim$\,$5$).
It constitutes perhaps the paradigm of a very massive star undergoing the
transition phase from the Main Sequence towards the Wolf-Rayet stage.
Some eruptive phenomena occurred in the past have produced spectacular
brightenings and mass ejections as those observed around 1843, when
the surrounding bipolar nebula known as ``Homunculus'' (Gaviola,~1950)
was expelled.

Nowadays, it is widely accepted that the object is made up of at least
a binary system, as first suggested by Damineli~(1996) and proposed
by Damineli et al.~(1997) based on the 5.5-year periodicity found for the optical
``spectroscopic events''. This hypothesis was reinforced by some other
associated phenomena registered in different wavelengths as the X-rays deep
minima (Corcoran,~2005), or the ``eclipse-like'' events in the optical bands
(van Genderen et al.,~2003; Fern\'andez-Laj\'us et al.,~2003) and near-infrared 
$JHKL$ photometry (Whitelock et al.,~2004).
The events were also reported at 7-mm (Abraham et al.,~2005) and 3-cm wave emission
(Duncan \& White,~2003).

The events would originate when the binary components are close to periastron,
and different radiative and wind interaction processes take place.
The last event at $\eta$~Car recently happened in January 2009, 
as predicted (e.g. Damineli et al.,~2008). 
This is the so-called 2009.0 event and points out
the beginning of ``cycle 12'' (we follow the nomenclature adopted by 
Groh and Damineli~(2004) to establish a common language to designate 
each event).
$BVRI$ optical photometry of a whole cycle was recently published by
Fern\'andez-Laj\'us et al.~(2009, hereafter PAPER I), as a result of a 
long-term monitoring campaign, which included the 2003.5 event.
Continuing with this monitoring we enlarged our observational 
capabilities including another telescope and filters, in order to 
guarantee complete and well sampled light curves of the 2009.0 event.

Our results are presented in this work to make them available for 
immediate modeling, analysis and interpretation.
A new estimation of the period length of ``cycle 11'' is also evaluated.

\section{Observations}
The 2009.0 event was monitored from two different observatories:
La Plata Observatory (OALP)\footnote{OALP belongs to Facultad de Ciencias
Astron\'omicas y Geof\'isicas, Universidad Nacional de La Plata 
(FCAG-UNLP), Argentina},
Buenos Aires, Argentina and Complejo Astron\'omico El Leoncito
(CASLEO)\footnote{CASLEO is operated under agreement between CONICET,
and the Universities of La Plata, C\'ordoba, and San Juan, Argentina},
San Juan, Argentina.
Our observing program involved the daily acquisition of CCD digital
images during the dates reported. Several series of 15 or 20 images, 
spanned no longer than 30 minutes each, were obtained for each filter 
every night. Three or four image series were typically acquired.
Observations started in early November 2008, after the annual 
visibility gap of the object.

\subsection{Image acquisition from OALP}
Observations were performed as a part of our 2009 observing season
(c.f. PAPER I), i.e. from November 12, 2008 (JDN 2454782) up to
March 30, 2009 (JDN 2454920), using the ``Virpi S. Niemela'' (VSN)
telescope as well as the same instrumentation and procedure 
described in PAPER I.
Besides the $BVRI$ filter set, a narrow passband (4.5 nm) H$\alpha$
filter was incorporated to the monitoring on December 26, 2008 (JDN 2454827).
The transmission peak of this filter is tuned at 656.3 nm.
A total of 14575 images were acquired during this period.

In addition to these data, this work presents another H$\alpha$ dataset 
acquired during the second half of our 2007 observing season and during 
the 2008 season. The H$\alpha$ filter used in this case is 6 nm-wide 
and is also tuned at 656.3 nm.

\subsection{Image acquisition from CASLEO}
CCD image acquisition was performed during November 27-29, 2008,
December 17-23, 2008, and January 07 to February 03, 2009, using a
Photometrics CH250 camera with a PM512 CCD chip attached to the
0.6 m Helen Sawyer Hogg telescope (f/15 Cassegrain) at CASLEO.
The chip array is 516\,x\,516 pixels (20 $\mu$m square pixel), giving
3$.\!{\!^{'}}$9 x 3$.\!{\!^{'}}$9 field images at the telescope focal plane.
In order to increase the signal-to-noise ratio, a 2\,x\,2 on-chip binning
factor was applied, the scale resulting 0$.\!{\!^{''}}$90 per pixel.
The image field was selected in such a way so as to enclose the 
field obtained at the VSN telescope.
About 10000 images were taken using a standard Johnson-Cousins $UBV$
filter set. 
Due to technical problems with the CCD camera cooling system, the camera 
had to be replaced by a similar one (with the same specifications mentioned 
above), on December 18, 2008. 
Therefore, the images taken since JDN 2454819 
were acquired with another camera unit, so these data could be affected 
by slightly different zeropoints.

\subsection{Data reduction}
The data reduction process was exactly the same as the one detailed in PAPER I,
HDE\,303308 being the comparison star for the $\eta$~Car relative photometry.
The aperture radius for extracting the instrumental magnitudes in the
$U,B,V,R$, and $I$ bands was again $12''$, enclosing the complete Homunculus.
For the H$\alpha$ photometry, a rather small aperture radius $= 3''$ was
used to get a higher signal-to-noise relation for the comparison star,
taking into account that this object is more than 6 magnitudes 
fainter than $\eta$~Car in this band.

The averaged errors of our differential photometry from OALP are:
$\epsilon_B=\pm 0.008$,
$\epsilon_V=\pm 0.005$,
$\epsilon_R=\pm 0.009$,
$\epsilon_I=\pm 0.013$, and
$\epsilon_{\rm{H}\alpha}=\pm 0.012$ mag;
and those from CASLEO are:
$\epsilon_U=\pm 0.005$,
$\epsilon_B=\pm 0.008$, and
$\epsilon_V=\pm 0.015$ mag.

\section{Results}
\begin{figure*}[!t]
\centering
        {\includegraphics[width=14.0cm,angle=0]{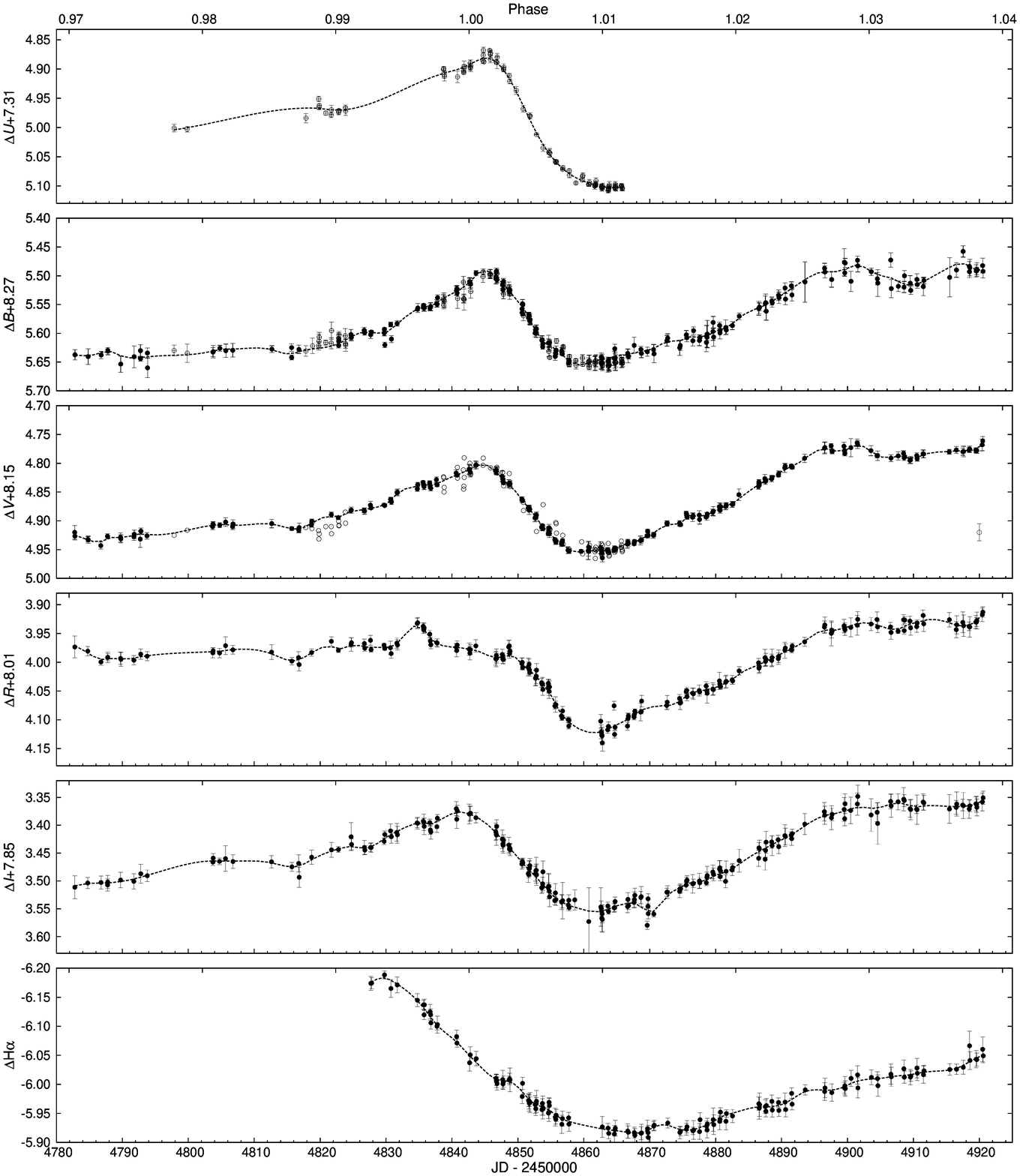}}
        \caption{
        $UBVRI$ and $H\alpha$ light curves of the 2009.0 ``eclipse-like'' event
        of $\eta$~Car, observed between November 12, 2008 and March 30, 2009.
        The $UBVRI$ zeropoints of HDE\,303308 were taken from Feinstein~(1982).
        Filled circles are data from OALP and open circles are those from CASLEO.
        Bars represent the standard deviations of the mean values. The typical error
        of the CASLEO $V$ data is represented separately (bottom right) for clarity.
        The OALP data were folded using smoothing splines as it was the $U$ light
        curve from CASLEO.
        Orbital phases are indicated at the top axis, according to the ephemeris
        given in Eq.~\ref{efem}.
        }
        \label{plots}
\end{figure*}

The resulting $UBVRI$ and H$\alpha$ light curves of the 2009.0 ``eclipse-like''
event are depicted in Fig.~\ref{plots}.
In this figure, we have used as zeropoints the $UBVRI$ Johnson-Kron-Cousins
photometry of HDE303308 provided by Feinstein~(1982), namely $U = 7.31$,
$B = 8.27, V = 8.15, R = 8.01$, and $I = 7.85$. No zeropoint was applied
to the H$\alpha$ data.
In order to match the $B$ and $V$ light curves of CASLEO data with those of
the OALP, we added $0.02$ mag ($0.05$ if JDN $< 2454819$) to the CASLEO $B$ 
data and $-0.09$ mag ($-0.11$ if JDN $< 2454819$) to the $V$ data.
The top axis depicts the orbital phase ($\phi$) derived from the ephemeris
\begin{equation}
        JD (\phi = 0) = \rm 2452819.8 + 2022.\!{\!^d}8 \cdot {\it E}\\
        \label{efem}
\end{equation}
where the date of $\phi = 0$ is given by Damineli et al.~(2008) and the period
is that derived in section 4.

%----------------------------------------------------------- 

%_____________________________________________________________

The OALP $BVRI$ light curves consist of about 200 data points each, and 138
for H$\alpha$. All of them are folded using natural cubic splines.
About 100 data points constitute the $UBV$ photometry from CASLEO, and the
$U$ curve was also smoothed using splines.
These data are available as an electronic table at the CDS
\footnote{Centre de Donn\'ees astronomiques de Strasbourg,
http://webviz.u-strasbg.fr/viz-bin/VizieR}.
An example of such table is shown in Table 2 in PAPER I.

\begin{figure}
\centering
        {\includegraphics[width=9.0cm,angle=0]{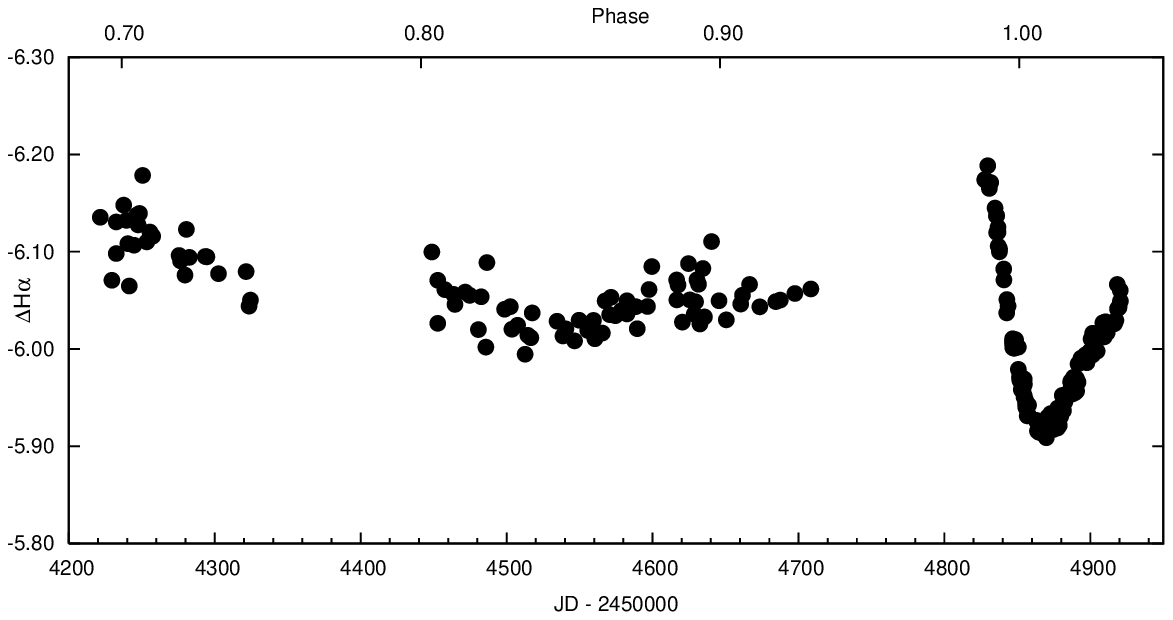}}
        \caption{
        H$\alpha$ light curve presented in Fig.~\ref{plots} including also 
	the data of the 2007 and 2008 observing seasons taken from OALP. 
	They were obtained using two different H$\alpha$ filters.
        Orbital phases are indicated at the top axis.
        }
        \label{halfa}
\end{figure}

The light curves are featured by an ascending branch starting at about
JDN 2454816 and lasting almost 30 days, when a maximum is reached.
This maximum peaks at different dates depending on the photometric 
band and is followed by a steep fading towards the minimum.
In the $R$ band the lack of the first ascending branch produces no
maximum, with the exception of a sharp peak centered at JDN 2454834.
The $U$ light curve is not complete due to the unavailability of
observing time at CASLEO, but the general behaviour is evident.
Unfortunately, the H$\alpha$ line photometry was not recorded during
the event before JDN 2454827 when the maximum apparently occurred.
Although the H$\alpha$ data of the 2007 and 2008 observing seasons
have higher rms than the 2009 data, the light curves during 
these seasons are quite constant and fairly fainter than before 2009,
highlighting the variations in the 2009.0 event, as seen in Fig.~\ref{halfa}.
%----------------------------------------------------------- S_vib
%_____________________________________________________________

{The maxima and minima occurrence dates are presented in Table~\ref{minmax}.
By comparing this monitoring with others}, we found that the starting 
of the minimum registered in the $V$ band takes place 15 days after
the phase of minimum excitation (Damineli et al.,~2009) and 11 days
after the starting of the minimum in X-rays (Corcoran,~2009).
This is in good agreement with what happened in the 2003.5 event
(see PAPER I).

%_____________________________________________________________
\begin{table}
\caption{
Julian Day Numbers for the brightness maxima and minima, and 
dip depths from our $UBVRI$ H$\alpha$ differential photometry of $\eta$~Car.
}
\label{minmax}
\centering
\begin{tabular}{c | c | c | c}
\hline\hline
Band &Maximum   & Minimum & Depth [mag] \\
\hline
$U$  & 2454845                  & 2454863  & 0.23       \\
$B$  & 2454845                  & 2454859  & 0.16       \\
$V$  & 2454844                  & 2454858  & 0.15       \\
$R$  & 2454834$^{\dag}$         & 2454861  & 0.15-0.18  \\
$I$  & 2454841                  & 2454861  & 0.18       \\
H$\alpha$ &  2454829 ?          & 2454862  & 0.26       \\
\hline
\noalign{\smallskip}
\multicolumn{4}{l}{\small (\dag) Peak time}\\
\end{tabular}
\end{table}
%_____________________________________________________________

After the minimum, a second ascending branch develops until
almost the same brightness reached at maximum is recovered.
The H$\alpha$ light curve shows a less pronounced rising tilt and does not
recover the original level.

All these features resemble the behaviour registered during the 2003.5 event.
The main difference is that the 2009.0 minimum is deeper than the previous
one (about 0.02-0.03 mag in all bands, e.g. Fig.~\ref{period}), and the
recovering branch is also steeper.

\section{Period fitting}
\begin{figure}
\centering
        {\includegraphics[width=9.0cm,angle=0]{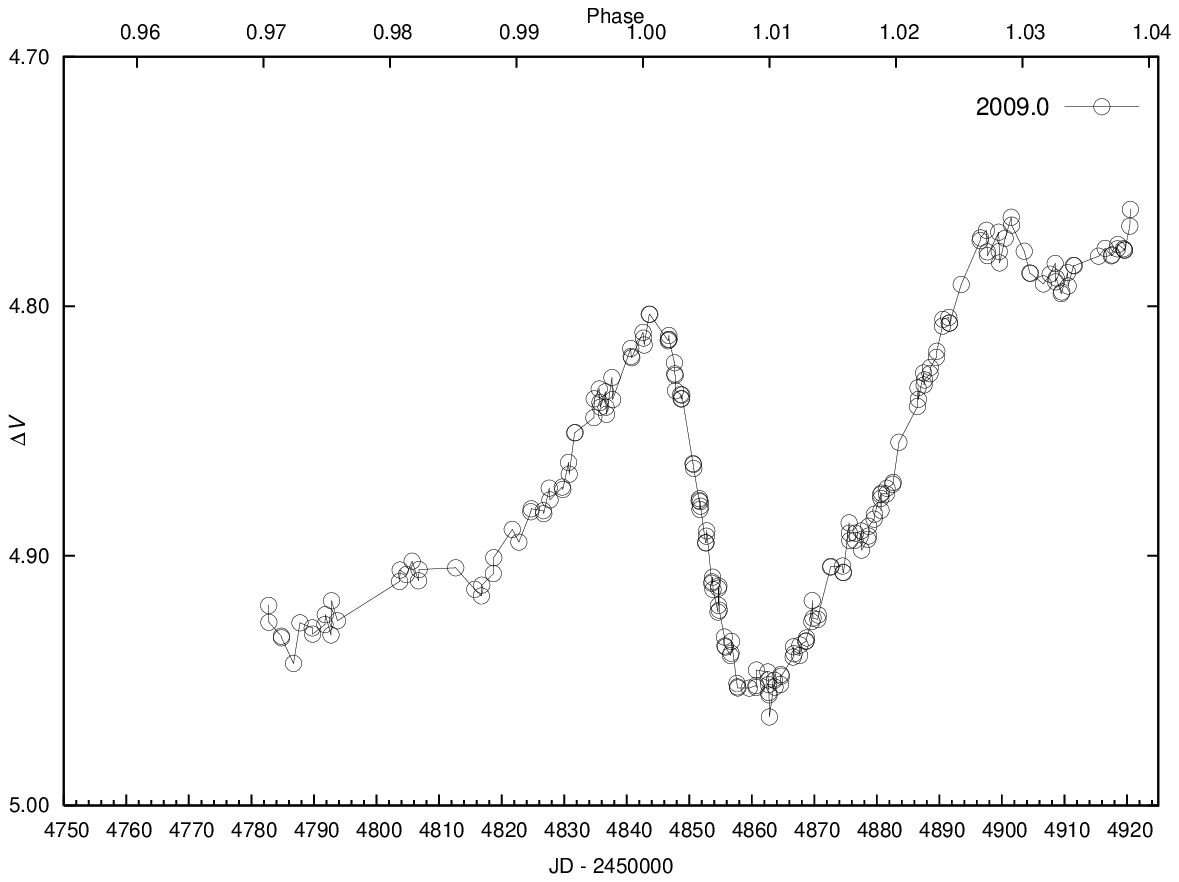}}
        \caption{
        $V$ light curves of the 2003.5 and 2009.0 ``eclipse-like'' events
        observed from OALP.
        Filled circles are data from 2009.0 and open circles are those from
        2003.5. We added -0.18 mag to the 2003.5 curve to match both curves.
        The data were folded using smoothing spline.
        Orbital phases are indicated at the top axis, according to the ephemeris
        given in Eq.~\ref{efem}.
        }
        \label{period}
\end{figure}

The time elapsed from one ``eclipse-like'' event to the next can be
used to determine the present value of the orbital period.
To inspect for any feature sharp enough to be used as a time 
reference point, we overlapped our 2003.5 dataset (PAPER I) with the 
new one.
It can be seen from the $V$ light curves of Fig.~\ref{period} that 
maxima or minima are not useful for this purpose.
The maxima are not resolved enough so as to match them and they 
seem to occur at slightly different phases.
To determine a minimum time is not possible either since both
minima are different.
Nevertheless, it would be a good alternative to consider the ascending
branch, the dip, and the first part of the recovering phase, i.e. from 
phase 0.986 up to phase 1.02 in Fig.~\ref{period}.
Using this range of data for each event, we attempted to match both
datasets by means of the ``phase dispersion minimization'' method
(Stellingwerf,~1978) for many trial periods.
Thus, we applied this method to each of our four $BVRI$ datasets from
OALP, testing periods ranging from 2018 to 2030 days,
with a step of 0.1 days.
The limits of this range of periods are the constraints obtained by 
Damineli et al. (2008) from the spectroscopic events.
As a result, we obtained that the most significant period values,
common to all photometric bands, span between 2021.9 and 2023.5 days,
2022.6 days being the mean value.

By applying this method again, using exclusively our OALP $V$ data 
of the 2003.5 and 2009.0 events, together with the $V$ data of the 
2003.5 event taken from Table 5 of van Genderen et al.~(2006), we 
found out that the most significant period values range between 
2022.8 and 2023.4 days.

Considering the set of periods yielded by the method, we resorted to 
a visual inspection to distinguish between possible and wrong values.
To such end, we phased all our $BVRI$ photometric datasets and checked
how every period matches both the 2003.5 and 2009.0 events.
We noticed that 2022.8 days fulfill this condition.
From the inspection, we also found that the values out of the range
2022.3-2023.3 days show a clear phase shifting between both events.
Therefore, we adopted for the period length of ``cycle 11'':
P = $2022.8 \pm 0.5$ days, which nicely matches all the 
observed bands.

%----------------------------------------------------------- S_vib
%______________________________________________________________

\section{Concluding remarks}
It was shown in this paper that the 2009.0-event optical photometric 
behaviour happened on schedule, as announced in PAPER I.
In the optical range the event exhibited once again an ``eclipse-like''
light curve. It was evident in all the $UBVRI$ bands and also in
the narrow H$\alpha$ band. In spite of this overall similar behaviour,
the time of occurrence of some features and other photometric details
(for instance the depth of the dips) differ in each band,
especially in $R$ and H$\alpha$.

The new estimation of the current period length is in complete
agreement with the average value derived by Damineli et al.~(2008)
from different spectral features and photometric bands.

The periodic recurrence of the observed events is verified by
the fact that this 2009.0 event occurred at the time predicted 
some time ago. This periodicity and the ``eclipse-like'' feature 
displayed in the optical light curves strongly support the proposal 
of the binary nature of $\eta$~Car.
Undoubtedly, further and exhaustive analysis with new information of
the 2009.0 event, coming from other wavelength ranges and spectral
features, will provide a better understanding of the intriguing
$\eta$~Car.

\vspace{0.8cm}\hspace{-17pt}{\bf Acknowledgements\/}\vspace{0.4cm}

The authors acknowledge the authorities of the FCAG-UNLP and CASLEO
for the use of their observational facilities.
We warmly thank the technical staffs of both observatories for the 
maintenance of and improvements to the telescopes and their equipments.
We acknowledge the participation of the following students of the
FCAG-UNLP during the observations: M. Haucke, C. Peri, and C. Scalia.
We are grateful to A. Cuestas and G. Bosch for the English review and
suggestions.
AFT acknowledges support by Agencia de Promoci\'on Cient\'ifica y 
Tecnol\'ogica with the grant PICT 111 BID 1728 OC/AR.
We thank the referee for his suggestions to improved the presentation 
of this paper.



\end{document}